\documentclass{PoS}

\title{e-MERLIN observations of the puzzling TeV source HESS J1943+213}

\ShortTitle{e-MERLIN observations of the puzzling TeV source HESS J1943+213}

\author{\speaker{Krisztina \'Eva Gab\'anyi}
\thanks{We thank the e-MERLIN staff for extensive support and help in data reduction. The e-MERLIN is a National Facility operated by the University of Manchester at Jodrell Bank Observatory on behalf of the UK Science and Technology Facilities Council (STFC). The research leading to these results has received funding from the European Commission Seventh Framework Programme (FP/2007-2013) under grant agreement No. 283393 (RadioNet3). We acknowledge support from the Hungarian Scientific Research Fund (OTKA K104539).}\\
        Departments of Theoretical and Experimental Physics, University of Szeged, Hungary\\
        F\"OMI Satellite Geodetic Observatory, Hungary\\
        Konkoly Observatory, MTA Research Centre for Astronomy and Earth Sciences, Hungary\\
        E-mail: \email{gabanyik@sgo.fomi.hu}}

\author{Gloria Dubner, Elsa Giacani\\
        Instituto de Astronom\'{\i}a y F\'{\i}sica del Espacio, Argentina\\
        }

\author{S\'andor Frey\\
        F\"OMI Satellite Geodetic Observatory, Hungary\\
        }

\author{Zsolt Paragi\\
        JIVE, Netherlands\\
        }

\abstract{HESS\,J1943+213 is a TeV source close to the Galactic plane proposed to be a BL Lac object. Our high resolution EVN observation failed to recover two thirds of the source flux density detected simultaneously by the WSRT. Our recent e-MERLIN observations in L and C bands show only a point source with flux density comparable to the EVN detection. Thus the structure responsible for the missing flux density has to be larger than $2^{\prime\prime}$. It may be related to the presumed extragalactic source (thus would have a kpc-scale size), or to the Galactic foreground material close to the line of sight to the source.}

\FullConference{12th European VLBI Network Symposium and Users Meeting \\
		7-10 October 2014\\
		Cagliari, Italy}

\begin{document}

\section{Introduction}
HESS\,J1943+213 is a TeV source very close to the Galactic Plane discovered by the H.E.S.S. Collaboration \cite{disc}. Counterparts of the source were identified in X-rays (CXOU\,J194356.2+211823), infrared (2MASS\,J19435624+2118233), and in radio (NVSS\,J194356+211826). 
In the discovery paper, three possible explanations were offered for the nature of the source: gamma-ray binary, pulsar wind nebula, or BL Lacertae object.  
Leahy and Tian \cite{tian}, based upon the HI spectrum towards the radio counterpart, proposed that the source is located beyond $16$\,kpc from the Sun, thus they argued for the extragalactic, BL Lac nature of the source. Recently, Peter et al. \cite{hostgalaxy} studied the near-infrared counterpart of the source in K band. The observations are consistent with the infrared source representing the host galaxy of HESS\,J1943+213, a typical massive elliptical galaxy. Additionally, the authors analyzed five years of {\it Fermi} LAT \cite{fermi} data and found a high-energy counterpart of the source. Using the infrared and gamma-ray data, they conclude that it is most likely located at a redshift between 0.03 and 0.45. Also recently, Tanaka et al. \cite{Fermi_source} presented new {\it Suzaku} \cite{Suzaku} X-ray observations. With a more conservative approach than \cite{hostgalaxy}, they derive only upper limit for the gamma-ray flux from the {\it Fermi} data. They conclude that these data, and the fact that the spectral energy distribution of HESS\,J1943+213 is analogous to other extreme high-peaked BL Lac objects, strongly indicate the BL Lac nature of the source.

We observed the source with the European VLBI Network (EVN) on 2011 May 18 \cite{sajat} at 1.6\,GHz and found a point-like counterpart at the position of the X-ray counterpart. (The radio position is $3.75^{\prime\prime}$ offset from the coordinates given in the NVSS catalog.) The flux density recovered in the EVN observation ($31\pm 3$\,mJy) was one-third of the flux density measured simultaneously with the Westerbork Synthesis Radio Telescope ($\approx95$\,mJy), which in turn is in good agreement with archival Very Large Array (VLA) C-configuration data ($91 \pm 5$\,mJy, beamsize: $18^{\prime\prime} \times 15^{\prime\prime}$) taken at 1.4\,GHz on 1985 September 30 (project: AH196). 


\section{Observations and data reduction}
To investigate the sub-arcsecond scale radio structure, we conducted e-Multiple Element Remotely Linked Interferometer Network (e-MERLIN) observations at 1.5\,GHz and at 5\,GHz (project: CY1017). The 1.5-GHz observation took place on 2013 December 7, and lasted for 12 h. The on-source time was $\approx6.0$ h (however, only $\approx 4.3$ hours could be used for the imaging of the source as for the first few hours the self-calibration did not provide meaningful solutions). The phase-reference calibrator was J1946+2300. 
In October 2013, the Jodrell Bank MkII telescope was undergoing maintenance, it could only participate in observations during the night. Therefore the 5-GHz observation was split into three 4.5-h long runs performed between 2013 October 11 and 14. Additional two runs were carried out on 2014 June 12 and 13 to cover the missing hour angles. Except for the last run, when MkII was not involved in the observation, all e-MERLIN telescopes participated. The on-source time was approximately $13$ h. The phase-reference calibrator was J1925+2106.

Data reduction was done using the National Radio Astronomy Observatory (NRAO) Astronomical Image Processing System (AIPS, \cite{aips}), following the e-MERLIN cookbook\footnote{version 2.4a, http://www.e-merlin.ac.uk/data\_red/tools/e-MERLIN\_cookbook\_latest.pdf}, and with extensive help from the e-MERLIN science team (P. E. Belles). 

\section{Results} 



\begin{figure}
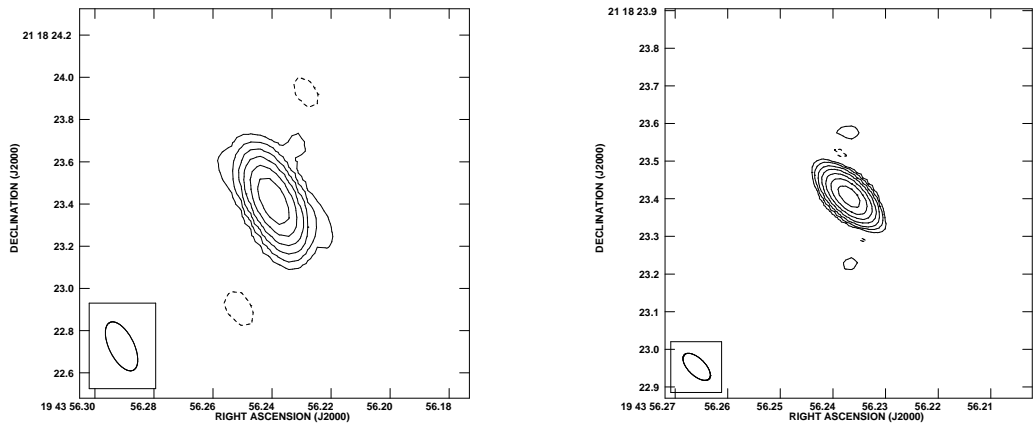

\begin{minipage}{0.5\textwidth}
\centering
\includegraphics[scale=0.3, angle=-90]{HESS_eMERLIN_L.ps}
\end{minipage}
\begin{minipage}{0.5\textwidth}
\centering
\includegraphics[scale=0.3, angle=-90]{HESS_eMERLIN_C2.ps}
\end{minipage}
\caption{e-MERLIN images of HESS\,J943+213. Left-hand side: L-band image. The peak is 19.2 mJy/beam, the beam size is 252\,mas $\times$ 117\,mas at a position angle of $26^\circ$. The lowest positive contour is at 0.34 mJy/beam ($7\sigma$-level), further contours increase with a factor of two. Right-hand side: C-band image. The peak is 19.4 mJy/beam, the beam size is 92\,mas $\times$ 45\,mas at a position angle of $46^\circ$. The lowest positive contour is at 0.2 mJy/beam ($7\sigma$-level), further contours increase with a factor of two.}
\label{eMERLIN}
\end{figure}

The e-MERLIN images are shown in Fig.\ref{eMERLIN}. The observations revealed an unresolved point source at both frequencies. We failed to detect any large-scale feature down to $7 \sigma$ level (0.34 \,mJy/beam) in a field of view of $23^{\prime\prime} \times 23^{\prime\prime}$ at L band. We used Difmap \cite{difmap} to fit brightness distribution models to the visibilities and found that the source can be best described with single circular Gaussian components at both frequencies. The emission has a flux density of $22.2 \pm 0.7$\,mJy, and $22.4\pm 0.3$\,mJy at 1.5 GHz and 5\,GHz, respectively. Thus, the observed feature is compact and has a flat-spectrum (but note that the L- and C-band observations were not simultaneous). The lower flux density at L band compared to the EVN value indicates changes in the brightness of this feature. The flux density variability and compactness of the detected feature is consistent with the proposed BL Lac nature of the source.
It seems that the large-scale radio emission detected with WSRT during our EVN run in 2011, and with the VLA in 1985, has been resolved out in the e-MERLIN observations. Thus $\approx60$\,mJy flux density is still missing compared to the WSRT and archival VLA observations. (Since the WSRT measurement was taken simultaneously with the EVN observation, source-intrinsic changes cannot explain the discrepant flux density values.) 

The large-scale radio structure seen in the archival VLA image has a size of $\approx1^\prime$. At the assumed redshift interval of the compact object, $0.03\le z \le 0.45$ \cite{hostgalaxy}, this size would correspond to 36 -- 349\,kpc (using standard $\Lambda$CDM cosmology with the parameters $H_0=67.3$\,km\,s$^{-1}$\,Mpc$^{-1}$, $\Omega_\textrm{m}=0.315$ \cite{planck}). The maximum angular scale that e-MERLIN can recover is $2^{\prime\prime}$ at L band. Thus the possible source of the missing flux density at 1.5\,GHz can only be in a structure with angular size larger than $2^{\prime\prime}$ (but smaller than $1^\prime$). At the assumed redshift range of the compact object, the lower size limit corresponds to approximately $1.3$\,kpc and $11.9$\,kpc linear extent. The existence of such extended features in BL Lac objects were deduced by Liuzzo et al. \cite{Marcello}, who found that most of their studied 42 low-redshift BL Lac objects have parsec-scale flux densities significantly lower than expected from the kiloparsec-scale flux densities measured by low-resolution surveys (such as the NVSS). 
On the other hand, the source is located at low Galactic latitude ($\approx-1.3^\circ$), the large-scale structure could also be related to Galactic material close to the line of sight towards the compact, presumably extragalactic source.

\section{Summary}
The most recent papers by Tanaka et al. \cite{Fermi_source} and Peter et al. \cite{hostgalaxy} suggest that the TeV source HESS\,J1943+213 is a BL Lac object. 
The source is point-like at the high resolution of EVN (10-mas scale) and also with the e-MERLIN (100-mas scale), according to our new L- and C-band data. The detected compact radio feature has a flat spectrum and in L-band showed flux density variability thus consistent with the proposed BL Lac nature of the source. Compared to the lower resolution VLA and WSRT observations at L band, significant flux density ($\approx 60$\,mJy) is missing from the high-resolution data. Whether this large-scale radio structure is related to the TeV-emitting BL Lac object or associated with the Galactic material in the line of sight towards the source is yet unclear. 



\end{document}